\documentstyle[12pt]{article}
\begin{document}
\title{{\bf Susskind's Challenge to the \\
Hartle-Hawking No-Boundary Proposal \\
and Possible Resolutions}
\thanks{Alberta-Thy-07-06, hep-th/0610199}}
\author{
Don N. Page
\thanks{Internet address:
don@phys.ualberta.ca}
\\
Institute for Theoretical Physics\\
Department of Physics, University of Alberta\\
Room 238 CEB, 11322 -- 89 Avenue\\
Edmonton, Alberta, Canada T6G 2G7
}
\date{(2006 December 15)}

\maketitle
\large
\begin{abstract}
\baselineskip 18 pt

	Given the observed cosmic acceleration, Leonard Susskind has
presented the following argument against the Hartle-Hawking
no-boundary proposal for the quantum state of the universe:  It should
most likely lead to a nearly empty large de Sitter universe, rather
than to early rapid inflation.  Even if one adds the condition of
observers, they are most likely to form by quantum fluctuations in de
Sitter and therefore not see the structure that we observe.  Here I
present my own amplified version of this argument and consider
possible resolutions, one of which seems to imply that inflation
expands the universe to be larger than $10^{10^{10^{122}}}$ Mpc.

\end{abstract}
\normalsize

\baselineskip 14 pt

\newpage

\section*{Introduction}

	Our part (or subuniverse \cite{Wein87} or bubble universe
\cite{Linde96,Vil97} or pocket universe \cite{Guth00}) of the entire
universe (or multiverse
\cite{James,Lodge,Leslie,Gell-Mann,Deutsch,Dyson,Rees,Carr} or
metauniverse \cite{Vil95} or omnium \cite{Penrose-book} or megaverse
\cite{Susskind-book} or holocosm \cite{Page-in-Carr}) is observed to
be highly special in a way that does not seem to be implied purely by
the known dynamical laws.  For example, it is seen to be very large on
the Planck scale, with low large-scale curvature, and with approximate
homogeneity and isotropy of the matter distribution on the largest
scales that we can see today.  It especially seems to have a very high
degree of order in the early universe that has enabled entropy to
increase, as described by the second law of thermodynamics
\cite{Tolman,Davies,Penrose-in-HI}.

	Two leading proposals for special quantum states of the
universe are the Hartle-Hawking `no-boundary' proposal
\cite{HVat,HH,H,HalHaw,P,Hal,HLL,HH1,Page-Hawking-BD,HH2} and the
`tunneling' proposal of Vilenkin, Linde, and others
\cite{V,TL,TZS,TR,TVV,TGV}. In toy models incorporating presumed
approximations for these proposals, both of these models have seemed
to lead to low-entropy early universes and so might explain the second
law of thermodynamics. If a suitable inflaton is present in the
effective low-energy dynamical theory, and if sufficient inflation
occurs, both proposals have seemed to lead to approximate homogeneity
and isotropy today.

	Here I shall focus on the Hartle-Hawking no-boundary proposal,
in which the wavefunction of the universe (expressed as a functional
of the 3-dimensional geometry and matter field configuration on a
spatial 3-surface that in some sense represents the universe at any
one moment of time) is given by a path integral over all compact
Euclidean 4-dimensional geometries and matter fields that have the
3-dimensional geometry and matter field configuration as its only
boundary.  (Because of this one boundary of the 4-geometries, where
the wavefunction is evaluated, one might say that the proposal would
be better named the Hartle-Hawking `one-boundary' proposal
\cite{Page-Hawking-BD}, but here I shall continue to use the usual
nomenclature.)  This model is certainly incomplete for various
technical reasons \cite{GHP,GS,Deser,Page-Hawking-BD}, but in simple
toy models, it seems to predict several special features of the
observed universe
\cite{HH,H,HP,HL,HalHaw,Page85,HLL,Page-Hawking-BD}: Lorentzian
signature, large size, near-critical density, low anisotropies,
inhomogeneities starting in ground states to fit cosmic microwave
background radiation (CMB) data, and entropy starting low to explain
the second law of thermodynamics.

	However, Leonard Susskind \cite{Susspriv} (cf.\
\cite{DKS,GKS,Sus03}) has argued that the cosmological constant or
quintessence or dark energy that is the source of the present
observations of the cosmic acceleration
\cite{Perl,Riess,PTW,Tonry,WMAP,Tegmark,Astier} would give a large
Euclidean 4-hemisphere as an extremum of the Hartle-Hawking path
integral that would apparently swamp the extremum from rapid early
inflation.  Therefore, to very high probability, the present universe
should be very nearly empty de Sitter spacetime, which is certainly
not what we observe.  

	This argument is a variant of Vilenkin's old objection
\cite{VvsHH} that the no-boundary proposal favors a small amount of
inflation, whereas the tunneling wavefunction favors a large amount. 
Other papers that have attacked the Hartle-Hawking wavefunction
include \cite{BC,GrT,BP}.  However, Susskind was the first to impress
upon me the challenge to the Hartle-Hawking no-boundary proposal from
the recent cosmic acceleration.

	Of course, it may be pointed out that most of de Sitter
spacetime would not have observers and so would not be observed at
all, so just the fact that such an unobserved universe dominates the
path integral is not necessarily contrary to what we do observe.  To
make observations, we are restricted to the parts of the universe
which have observers.  One should not just take the bare probabilities
for various configurations (such as empty de Sitter spacetime in
comparison with a spacetime that might arise from a period of rapid
early inflation).  Rather, one should consider conditional
probabilities of what observers would see, conditional upon their
existence \cite{Vil95,Page-con,HH2}.

	However, the bare probability of an empty de Sitter spacetime
forming by a large 4-hemisphere extremum of the Hartle-Hawking path
integral dominates so strongly over that of a spacetime with an early
period of rapid inflation that even when one includes the factor of
the tiny conditional probability for an observer to appear by a vacuum
fluctuation in empty de Sitter, the joint probability for that
fluctuation in de Sitter dominates over the probability to form an
inflationary universe and thereafter observers by the usual
evolutionary means.  Therefore, the argument goes, almost all
observers will be formed by fluctuations in nearly empty de Sitter,
rather than by the processes that we think occurred in our apparently
inflationary universe.

	The problem then is that almost all of these
fluctuation-observers will not see any significant ordered structures
around them, such as the ordered large-scale universe we observe.
Thus our actual observations would be highly atypical in this
no-boundary wavefunction, counting as strong observational evidence
against this theory (if the calculation of these probabilities has
indeed been done correctly).  As Dyson, Kleban, and Susskind put it
in a more general challenge to theories with a cosmological constant
\cite{DKS}, ``The danger is that there are too many possibilities
which are anthropically acceptable, but not like our universe.''  See
\cite{decay,BF,LindeBB,return,Vilfreak,astronomical} for further
descriptions of this general problem.

	To express this in a slightly different way, if $A$ are the
conditions for observations, and if $B$ are the conditions for ordered
observations, we want a theory giving the conditional probability
$P(B|A)$ not too many orders of magnitude smaller than unity, since we
see $B$.  But if the no-boundary quantum state produces $A$ mostly by
de Sitter fluctuations, it seems that it gives $P(B|A) \ll 1$.

	The general nature of this objection was forcefully expressed
by Eddington 75 years ago \cite{Edd}:  ``The {\it crude} assertion
would be that (unless we admit something which is not chance in the
architecture of the universe) it is practically certain that at any
assigned date the universe will be almost in the state of maximum
disorganization.  The {\it amended} assertion is that (unless we admit
something which is not chance in the architecture of the universe) it
is practically certain that a universe containing mathematical
physicists will at any assigned date be in the state of maximum
disorganization which is not inconsistent with the existence of such
creatures.  I think it is quite clear that neither the original nor
the amended version applies.  We are thus driven to admit anti-chance;
and apparently the best thing we can do with it is to sweep it up into
a heap at the beginning of time.''

	In Eddington's language, Susskind's challenge is that the
Hartle-Hawking no-boundary proposal seems to lead to pure chance (the
high-entropy nearly-empty de Sitter spacetime), whereas to meet the
challenge, we need to show instead that somehow in the very early
universe (near, if not at, the ``beginning of time'') it actually
leads to anti-chance, something far from a maximal entropy state.

	Of course, another possibility is simply that the
Hartle-Hawking no-boundary proposal is wrong.  Hawking himself
admitted this possibility \cite{Hawk-brief} (cf. also \cite{HH2}): 
``I'd like to emphasize that this idea that time and space should be
finite without boundary is just a {\it proposal}: it cannot be deduced
from some other principle.  Like any other scientific theory, it may
initially be put forward for aesthetic or metaphysical reasons, but
the real test is whether it makes predictions that agree with
observation.''  Susskind is making the argument that its predictions
do not agree with observation.

\section*{Numerical Illustrations}

	Let us make a numerical illustration of this problem.  For
simplicity and concreteness, let us take $\Omega_\Lambda = 0.72\pm
0.04$ from the third-year WMAP results of \cite{WMAP} and $H_0 =
72\pm 8$ km/s/Mpc from the Hubble Space Telescope key project
\cite{Freedman}, and let us drop the error uncertainties.  In Planck
units, $\hbar = c = G = 1$, this gives $H_0 \approx 1.3\times
10^{-61}$ and $\Lambda = 3\Omega_\lambda H_0^2 \approx 3.4\times
10^{-122}$, which would give a Euclidean 4-hemisphere of radius
$a_{\mathrm{dS}} = \sqrt{3/\Lambda} \approx 9.4\times 10^{60}$ and a
Euclidean action of $S_E(\mathrm{de\ Sitter}) = -\pi
a_{\mathrm{dS}}^2/2 \approx -1.4\times 10^{122}$.  This extremum of
the Hartle-Hawking path integral would thus give an unnormalized bare
probability of
\begin{equation}
P_{\rm bare}(\mathrm{de\ Sitter}) = e^{-2S_E(\mathrm{dS})}
 = e^{\pi a_{\mathrm{dS}}^2} = e^{3\pi/\Lambda} \sim e^{10^{122.44}}.
\label{eq:1}
\end{equation}

	Now we need to ask, given this de Sitter spacetime, what is
the probability of having an observer or observation.  I do not know
what the minimum requirement for an observation is, but it certainly
seems sufficient to have a human brain in the right state for a
sufficient time.  If we assume that a human brain of minimum mass,
say, 1 kg, can make an observation in a very short time if it is in
the right state, then the minimum requirement would be for the brain
to fluctuate into existence in a region of size, say, $r=$ 30 cm,
that is separate from the antimatter that would also exist during the
vacuum fluctuation.  This gives a dimensionless action of (cf.\
\cite{42})
\begin{eqnarray}
S_E({\mathrm brief\ brain}) \sim {E r\over\hbar c}
\sim {(1\:\mathrm{kg})(3\times 10^8\:\mathrm{m/s})(0.3\:\mathrm{m})
\over 10^{-34}\:\mathrm{J}\cdot\mathrm{s}} \sim 10^{42}.
\label{eq:2a}
\end{eqnarray}

	This will then give a conditional probability of each brain
state, in some region of de Sitter spacetime just large enough to
contain the brain, given that that spacetime exists, of about
$e^{-2S_E(\mathrm{brain})} \sim e^{-10^{42}}$. This should be
multiplied by the number of orthogonal brain states within the region
that would correspond to an observation, and by the number of
spacetime regions where the observer can fluctuate into existence and
make an observation, in order to give $P(\mathrm{brain\ state}|
\mathrm{de\ Sitter})$.  However, for simplicity let us assume that the
product of these numbers is much less than $e^{10^{42}}$ and so does
not much change the upper exponent (42).

	A more conservative assumption \cite{Page-lifetime} would be
that a human observer requires a 1 kg brain to last a time long
enough for neural signals to travel across it, say 0.1 second.  The
dimensionless action for this is
\begin{eqnarray}
S_E({\mathrm medium\ brain}) \sim {E\Delta t\over\hbar}
\sim {(1\:\mathrm{kg})(3\times 10^8\:\mathrm{m/s})^2(0.1\:\mathrm{s})
\over 10^{-34}\:\mathrm{J}\cdot\mathrm{s}} \sim 10^{50}.
\label{eq:2b}
\end{eqnarray}

	An even more conservative assumption would be that the brain
should be a thermal fluctuation into a real existence from the de
Sitter temperature $T_{\mathrm{dS}} = 1/(2\pi a_{\mathrm{dS}}) =
1.70\times 10^{-62}$, which gives a Boltzmann probability factor of\\
$\exp{(-2S_E({\mathrm long\ brain}))}$ with
\begin{eqnarray}
S_E({\mathrm long\ brain}) = {\pi a_{\mathrm{dS}} E\over\hbar c}
\approx 1.4\times 10^{69}.
\label{eq:2c}
\end{eqnarray}
Effectively this assumption was used in \cite{BF} to calculate the
time for a `Boltzmann brain' (BB) \cite{Bolt,Rees,AS} to appear in
the local viewpoint considered there, $t_{BB} \sim
\exp{(2S_E({\mathrm long\ brain}))}$; here I shall suggest that it
might be more realistic to use the smaller brief brain (bb) time
$t_{bb} \sim \exp{(2S_E({\mathrm brief\ brain}))} \sim e^{10^{42}}$
in that viewpoint.

	In \cite{Page-lifetime} I used $S_E({\mathrm medium\ brain})
\sim 10^{50}$ to estimate that if the de Sitter spacetime lasts
longer than about $10^{50} t_0 \sim 10^{60}$ years, then the
spacetime 4-volume would be so large that one would expect many
observers to fluctuate into existence in it, rather than having just
a very low probability per de Sitter spacetime.  An even more severe
problem seems to occur if the spacetime decays probabilistically with
a half-life greater than about 20 billion years, since then the
expectation value of the 4-volume per comoving 3-volume would diverge
in the future and produce an infinite number of fluctuation observers
or Boltzmann brains \cite{decay,return,astronomical}.  However, here
let us assume either that the de Sitter spacetime will not last so
long, or else that there are other ways of circumventing these
problems, such as the ones suggested in \cite{BF,LindeBB,Vilfreak}.

	Combining the unnormalized bare probability for nearly empty
de Sitter with the most conservative conditional probability for an
observation within such a de Sitter spacetime gives the unnormalized
probability of an observation from the Euclidean de Sitter extremum
as
\begin{eqnarray}
P_{\rm observation, unnormalized}(\mathrm{de\ Sitter}) \sim
\exp{(+10^{122.44} - 10^{69.4})}.
\label{eq:3}
\end{eqnarray}

	In comparison, let us calculate the probability of forming an
observer through an inflationary universe.  In this case observers
can presumably develop through normal Lorentzian evolution (with
paths in the path integral having real Lorentzian action or purely
imaginary Euclidean action during the Lorentzian part of the
evolution) after one has Lorentzian inflation, so there will not be
the huge suppression factor of about $e^{-10^{42}}$, $e^{-10^{50}}$,
or $e^{-10^{69.4}}$ that occurs in empty de Sitter.  This by itself
certainly makes it sound as if more observers ought to be produced by
inflation than by empty de Sitter.  However, in the Euclidean path
integral of the Hartle-Hawking no-boundary proposal, one also needs
to compare the bare probability of producing the inflationary
universe, which seems to be much, much less than the bare probability
of producing a large de Sitter spacetime directly by the large
Euclidean 4-dimensional hemisphere extremum.

	Although the details are unimportant for the qualitative
result of the argument, for concreteness let us consider inflation
driven by a single scalar field $\phi$ with potential $V(\phi)$.  In
the Hartle-Hawking no-boundary path integral, inflation can start by
an extremum of the action that has a nearly-round Euclidean small
4-dimensional hemisphere with nearly constant scalar field value
$\phi_0$, radius squared $a_0^2 \approx 3/[8\pi V(\phi_0)]$, and
Euclidean action $S_E(\mathrm{inflation}) \approx -\pi a_0^2/2
\approx 3/[16 V(\phi_0)]$.

	In the account of Liddle and Lyth \cite{LL}, who use the
reduced Planck mass $M_{\mathrm Pl} = (8\pi G/\hbar c)^{-1/2} =
1/\sqrt{8\pi} \approx 0.20$ in terms of the usual Planck units
$\hbar=c=G=1$, the magnitude of the scalar density perturbations from
inflation is given by
\begin{eqnarray}
{\mathcal P_R}(k) = \left({H\over\dot{\phi}}\right)^2
                   \left({H\over 2\pi}\right)^2
= {1\over 24\pi^2 M_{\mathrm Pl}^4}{V\over\epsilon}
= {8V\over 3\epsilon},
\label{eq:4}
\end{eqnarray}
where
\begin{equation}
\epsilon \equiv \epsilon(\phi) 
= {M_{\mathrm Pl}^2\over 2}\left({V'\over V}\right)^2
\equiv {1\over 16\pi}
       \left({1\over V}{dV\over d\phi}\right)^2
\label{eq:5}
\end{equation}
is one of the slow-roll parameters \cite{LL} that I am assuming is
much less than unity, and everything is to be evaluated at horizon
exit for wavenumber $k = aH$.

	The Liddle-Lyth quantity ${\mathcal P_R}(k)$ seems to be the
same quantity called $\Delta_{\mathcal R}(k)$, the amplitude of
curvature perturbations, in the WMAP analysis \cite{WMAP}, which at a
wavenumber of $k = 0.002/{\mathrm Mpc}$ is given in terms of the
amplitude of density fluctuations, $A$, as $29.5\times 10^{-10} A$. 
Table 5 of the 3-year WMAP data \cite{WMAP} gives $A \approx 0.8$, so
I shall take that as a representative value below.

	Now if for simplicity and concreteness we suppose that the
inflaton potential has a power-law form with exponent $\alpha$, say
$V = \lambda M_{\mathrm Pl}^{4-\alpha} \phi^\alpha$, then the
slow-roll parameter is $\epsilon = (1/2)\alpha^2 M_{\mathrm
Pl}^2/\phi^2 \approx \alpha/(4N)$, where $N$ is the number of e-folds
of inflation from that value of $\phi$ to the end of inflation at
$\phi_{\mathrm end} \approx \alpha M_{\mathrm Pl}/\sqrt{2}$, assuming
that $\phi \gg \phi_{\mathrm end}$.

	In terms of these quantities, if the nearly-round Euclidean
4-hemisphere has the nearly-constant scalar field value $\phi_0$
(which can be interpreted to be very nearly the initial value for the
Lorentzian inflation that is the analytic continuation of the
Euclidean 4-hemisphere), and if $\phi_0$ is taken as a lower limit for
the value that causes the horizon exit at what is now the fiducial
wavenumber of $k = 0.002/{\mathrm Mpc}$, then the Euclidean action of
the 4-hemisphere is, roughly,
\begin{equation}
S_E \stackrel{>}{\sim} -{1\over 2\epsilon\Delta_{\mathcal R}}
\approx -{2N\over\alpha\Delta_{\mathcal R}}
= -6.78\times 10^8 {N\over\alpha A} \approx -2\times 10^{10},
\label{eq:6}
\end{equation}
where for the very last number I have taken $\alpha = 2$ for the
${1\over 2} m^2\phi^2$ potential that does seem to fit the WMAP data
\cite{WMAP} better than the $\lambda\phi^4$ potential with $\alpha =
4$, and I have used the value $N=50$ from \cite{LL} and the value
$A=0.8$ from \cite{WMAP}.  This result also agrees well with the
result of using $m = 7.5\times 10^{-6} = 1.5\times 10^{-6}$ and
$\phi_0 = \phi_* = 14 M_{\mathrm Pl} = 2.8$ for the ${1\over 2}
m^2\phi^2$ potential from the example on page 252 of \cite{LL}.  Very,
very crudely, when $\phi_0$ is not much larger than the value of
$\phi$ giving $k = 0.002/{\mathrm Mpc}$, then this Euclidean action
goes as the inverse square of a typical galactic peculiar velocity (in
units of the speed of light, of course), multiplied by $N$ that is
very roughly a logarithm of the ratio of some energy in the range of
the Planck energy to some energy in the range of atomic energies.

	Then if I use this estimate for a lower bound on the Euclidean
action for the 4-hemisphere, the unnormalized bare probability
for inflation becomes
\begin{equation}
P_{\rm bare}(\mathrm{inflation}) = e^{-2S_E(\mathrm{inflation})}
 = e^{\pi a_0^2} = e^{3/[8V(\phi_0)]}\stackrel{<}{\sim} e^{10^{10.6}}.
\label{eq:7}
\end{equation}

	If $\phi_0$ were taken to be larger, which is certainly
consistent with the observations that only place a minimum value on
the number of e-folds of inflation, then $P_{\rm
bare}(\mathrm{inflation})$ would be smaller, asymptotically
approaching unity as $V(\phi_0)$ approaches infinity with the
ever-rising potential.

	If inflation does occur, then one would expect the
conditional probability of observers to be of the order of unity (not
suppressed by a Euclidean fluctuation action), so one would get
\begin{eqnarray}
P_{\rm observation, unnormalized}(\mathrm{inflation})
\stackrel{<}{\sim} e^{10^{10.6}}.
\label{eq:8}
\end{eqnarray}
This is much less than $P_{\rm observation, unnormalized}(\mathrm{de\
Sitter})$, so if we normalize be dividing by the total unnormalized
probability for observations, we get that the normalized probability
for an observation to occur in an inflationary solution (rather than
from a fluctuation in nearly empty de Sitter) would be
\begin{eqnarray}
P_{\rm observation}(\mathrm{inflation}) \stackrel{<}{\sim}
 \exp{\left(-10^{122.44} + 10^{69.4} + 10^{10.6}\right)}.
\label{eq:9}
\end{eqnarray}

	In fact, if one just asks for the normalized probability of
an ordered observation, that would much more likely occur from a
fluctuation in the large nearly-empty de Sitter than in an
inflationary universe, so
\begin{eqnarray}
P_{\rm observation}(\mathrm{order})
\sim {\mathrm{number\ of\ brain\ states\ with\ ordered\ observations}
 \over\mathrm{number\ of\ brain\ states\ with\ any\ observations}}.
\label{eq:10}
\end{eqnarray}

	This would still be expected to be a fraction much less than
unity, depending on how ordered the observation is required to be. 
(For example, one might expect that the fraction of observations with,
say, 1000 ordered bits of information would be of the order of
$2^{-1000}$).  Therefore, according to these probabilities given by
this approximate calculation in a toy
minisuperspace-plus-homogeneous-inflaton model, it would be very
improbable for an observation chosen randomly from the predictions of
the model to have the order that we see in our actual observations. 
That is, our actual observations would be highly atypically ordered
according to this model, and this fact counts as strong observational
evidence against the model.

	The conclusion of Susskind's argument \cite{Susspriv}, which
I have expanded in my own words here, is that the Hartle-Hawking
no-boundary proposal for the quantum state of the universe is
inconsistent with our observations.

	I indeed take this as a very serious objection to the
no-boundary proposal, for which I do have not seen or thought of a
rebuttal that I would regard as completely satisfactory.  However,
since this proposal has in the past (at least in highly approximate
toy models) seemed to provide solutions for a number of deep cosmic
mysteries (perhaps foremost the explanation of the very low entropy of
the very early universe necessary to explain the second law of
thermodynamics), I am loathe to give it up.  Therefore, I would like
to regard Susskind's objection not so much as a no-go theorem but more
as a challenge (to discover either how to save the Hartle-Hawking
proposal or how to replace it).

\section*{Possible Resolutions}

	In this spirit, let me consider various possible resolutions
to Susskind's challenge to the Hartle-Hawking no-boundary proposal,
though readily admitting that none I have thought of yet seems to be
completely satisfactory.

	(1) The first conceivable resolution to Susskind's challenge
is that for some unknown reason observers can't form from fluctuations
in nearly-empty de Sitter spacetime (or for some reason they have
probabilities suppressed enormously much more greatly than that
calculated above for a brain to last 0.1 seconds).

	A separate motivation for this possibility is the calculation
of \cite{Page-lifetime} that if our current accelerating universe
lasts longer than about $10^{50} t_0 \sim 10^{111} \sim 10^{60}$
years into the future, the comoving 4-volume corresponding to the
Solar System, say, would have far more observers produced by vacuum
fluctuations than are likely to exist from ordinary life on Earth
over the entire history of the Solar System.  The results of
\cite{decay,return,astronomical} imply that this problem could arise
from a quantum half-life as low as 20 billion years (rather than an
end to the universe at a definite time that could be as long as $\sim
10^{60}$ years in the future \cite{Page-lifetime}).  Therefore, even
if we exclude nearly-empty de Sitter spacetimes formed in the
Hartle-Hawking path integral, it would seem that our ordered
observations would be highly atypical even within what we think is
happening in our part of the universe, if it lasts long enough.

	Of course, one possibility is that our part of the universe
does not last this long (at least while expanding exponentially at
roughly the present rate) \cite{Linde,Star,KLP,KL,ASS,KKLLS,GLV,WKLS,
Per,Page-lifetime,decay,return,astronomical}.  Perhaps the current
slow cosmic acceleration is not caused by a cosmological constant (or
by a scalar field at a positive minimum of its potential, which is
effectively essentially the same thing if the tunneling rate out from
this minimum is negligible in $10^{60}$ years).  Perhaps instead it
is due to a scalar field that is slowly sliding down a potential with
a slope \cite{Linde,Star,KLP,KL,ASS,KKLLS,GLV,WKLS,Per,Page-lifetime}
that is very small but which is sufficient for the potential to go
negative and lead to a big crunch within $10^{60}$ years.  (The
observational evidence presented in \cite{Page-lifetime} only gave a
lower limit of about 26 billion years in the future, assuming that
the potential is not convex.  Improved measurements of the $w$
parameter are expected to give a gradual improvement of this lower
limit, but it seems totally unrealistic to expect the observational
lower limit to be raised to $10^{60}$ years within the near future,
by which of course I mean within some humanly accessible time scale
$\ll 10^{60}$ years.)

	However, it would seem to require extraordinary fine tuning to
have a potential with a small enough nonzero slope to be consistent
with our observations and yet allow the universe to slide into
oblivion.  Having a minimum of the potential of low enough value to
give the observed cosmic expansion might be explained by the anthropic
principle (restricting attention to probabilities conditional upon
observers) \cite{Weinberg,MSW}, but there does not seem to be any
obvious similar argument why the current very low value (in Planck
units) of the potential should also be accompanied by a very low
nonzero value of the slope.  It would seem much more likely that one
were at a minimum of a potential than that one were at a low value of
a potential that also has a gradual slope extending far enough to lead
to negative values for the potential (and hence an eventual big crunch
for the universe).  Therefore, one is tempted to look for other
resolutions of the problem posed by the possibility of the production
of observers by vacuum fluctuations.

	One possibility for this is that observers require nonzero
globally conserved quantities that are almost entirely absent in
nearly-empty de Sitter spacetime.  However, this would seem to
require that observers must extend over the entire space, or else one
could simply have a fluctuation in which the required value of the
conserved quantity appeared in the smaller region where the observer
is, and then the complementary region would have the negative of this
quantity, so that the total quantity over the entire space remains
zero \cite{Page-lifetime}.  It seems rather implausible to propose
that as observers, each of us extends over all of space, though one
might note that similarly counter-intuitive things seem to occur in
the representation of a bulk gravitational quantum state on the
conformal boundary in the AdS/CFT correspondence.

	Therefore, I cannot rule out the possibility that
nearly-empty de Sitter space cannot produce observers by vacuum
fluctuations, but it does seem rather far-fetched to me to suppose
that it cannot.

	(2) A second conceivable resolution of Susskind's challenge is
that the Euclidean action of the inflationary universe can be made
very large and negative by connecting it by a thin bridge or tube or
thread to a large Euclidean de Sitter 4-sphere, thereby making its
Euclidean action even more negative than that of pure Euclidean de
Sitter without inflation \cite{Hartle-priv,HH2}.  However, then the
question would be that if this were possible, why not also have the
nearly-empty de Sitter itself also connected by a bridge to another
large 4-sphere to reduce its action as well and keep it more negative
than that of the inflationary universe?  Furthermore, if one allowed
one bridge to another 4-space of negative action, what prevents there
from having an arbitrarily large number of bridges connecting to an
arbitrarily large number of 4-spaces of negative action, thereby
making the Euclidean action unbounded below?  This would then seem to
make the theory degenerate into nonsense.

	So for this second possibility to be valid, it would seem
that there must be some unknown principle that allows an inflationary
universe to be connected to a large Euclidean de Sitter 4-sphere,
but not for the nearly-empty de Sitter 4-hemisphere to be similarly
connected to something else to reduce its action similarly.

	One proposal that might be sufficient to rule out the
catastrophe of an arbitrary number of bridges connecting some space in
the path integral to an arbitrary number of 4-spaces of negative
action would be that one should approximate the Euclidean path
integral by a sum only over actual extrema of the action, real or
complex Euclidean solutions of the Einstein equations coupled to the
matter fields \cite{Page-Hawking-BD}.  It would seem likely that this
would allow either of the two extrema discussed above, the large
4-dimensional hemisphere (which analytically extends into the
Lorentzian regime as nearly-empty de Sitter spacetime) and the
Euclidean inflationary solution (with its tiny approximately round
4-dimensional hemisphere followed by its analytical extension to a
Lorentzian inflationary universe), but perhaps not solutions with an
arbitrary number of bridges connecting different large Euclidean
regions with large negative action.  (When one considers the amplitude
for observers, one could still take just complex classical solutions,
but now slightly inhomogeneous ones that end up with the perturbed
final 3-space having an observer configuration as a fluctuation that
would raise the real part of the Euclidean action in the de Sitter
case but mainly just give an imaginary correction to the Euclidean
action in the inflationary case.)

	(3) This leads to the third conceivable resolution, which is
that there is an actual extremum connecting the inflationary solution
to a large negative-action Euclidean de Sitter, but none connecting
two Euclidean de Sitter spaces.  Then there would be a (probably
complex) Euclidean solution of the field equations with huge negative
action (from the Euclidean de Sitter part) and yet having a part that
analytically continues to a Lorentzian inflationary universe that can
explain our observations, without there being a solution with even
more negative Euclidean action (say from two Euclidean de Sitter
solutions somehow connected together to make a new solution with more
negative action).  Such an inflationary solution with a huge negative
action would almost necessarily be inhomogeneous, which might make it
difficult to discover.  There is no evidence that I am aware of that
strongly suggests its existence, but then there is none I know that
would rule it out either.

	This suggested resolution has the advantage that in principle
one could look for an explicit realization, though it might be
difficult.  The main trouble that I see with it is that it seems
somewhat implausible to me that an inflationary solution would have an
extension (a modified solution that includes a much larger complex
Euclidean region, not just an extension of the same solution), still
as a solution of the field equations, that includes a region giving
huge negative action, if one cannot do the same for the Euclidean de
Sitter solution.  And even if one somehow succeeded in finding such an
extension of the inflationary solution, it might be difficult to prove
that there is no analogous extension of the Euclidean de Sitter
solution.  Therefore, at present I would regard this conceivable
resolution as quite speculative, though it would be very exciting if
one could find an actual complex solution of the character outlined
above (with the appropriate nonsingular Euclidean boundary conditions
of the no-boundary proposal).  Such an actual mathematical solution
would be of greater scientific value than many of the speculative
proposals I am desperately tossing out for consideration in this
paper.

	(4) A fourth conceivable resolution, rather going in the
opposite direction from that of the previous one, is that Euclidean
de Sitter is not an allowed extremum of the path integral with a
cosmological constant.  It would seem likely that even if one
attempts to make the Hartle-Hawking path integral manageable by
restricting the sum to extrema, one might need to restrict the sum
only to a certain subset of all complex extrema.

	For example, it was found in one calculation \cite{HH} for a
3-dimensional sphere of size smaller than the equatorial 3-sphere of
the 4-sphere solution for the chosen value of the cosmological
constant, that even though there were two classical solutions (one in
which the 3-sphere boundary bounded less than half of the 4-sphere,
and the other in which it bounded more than half), only the solution
with the 3-sphere bounding less than half of the 4-sphere contributed
to a preferred contour integral for the path integral (and even though
the other solution had lower Euclidean action).

	It also might be expected that one could have complex
Euclidean solutions that wind around various singularities
\cite{Unruh-Jheeta,Page-Hawking-BD}, and that the real part of the
action could be made arbitrarily negative by winding around in the
appropriate direction.  In this case it would not seem to make sense
to include the solutions with arbitrarily negative action, so one
might need to make some restriction on the number of times the complex
solution could wind around various singularities.  However, it is not
clear to me what the correct procedure would be to accomplish this.

	Nevertheless, it might turn out that somehow the correct
procedure, once found, would rule out using the Euclidean de Sitter
extremum but would still allow the inflationary solution. Again, at
the moment this remains pure speculation, and it is hard to see why
something so simple as the Euclidean de Sitter 4-hemisphere (and its
analytic continuation into the Lorentzian regime) would be excluded.

	(5) A related fifth conceivable resolution is that even if
Euclidean de Sitter is an allowed solution that would contribute to
the Hartle-Hawking path integral with a true cosmological constant, it
(or a similar large 4-space) is not an allowed solution with the
actual quintessence or dark energy that drives the currently observed
cosmic acceleration.  It is hard to see how quintessence or dark
energy would not give a Euclidean solution if a cosmological constant
does, but I do not have a rigorous proof against this, so I am
therefore listing it as one of the conceivable possibilities.

	(6) A sixth possibility is that whatever resolves the problem
of the infinite measure in inflation might also in some way solve the
problem raised by Susskind (though I certainly don't see why this
would necessarily occur).  Inflation, particularly eternal inflation
\cite{Veternal,Leternal,Seternal,Geternal}, seems to be able to lead
to an arbitrarily large universe, with an arbitrarily large number of
observers, which makes it problematic how to calculate the
probability of various observed features by taking the ratio of the
numbers of the corresponding infinite sets of observers
\cite{LM,LLM,GL,Vil,WV,VVW,GV,GSVW,ELM,Bousso,BF,LindeBB,AGJ}.  It is
conceivable that the resolution of this dilemma might also regulate
the huge bare probability ascribed to the nearly-empty de Sitter
spacetime in the Hartle-Hawking path integral.  However, it is also
possible that the two problems are rather separate, so that a
solution to one will not immediately give a solution to the other.

	(7) A seventh possible resolution of Susskind's challenge is
that the integral over the initial value $\phi_0$ of the scalar field
$\phi$, being infinite if the $\phi$ has an infinite range, will
dominate over the huge but finite value of $P_{\rm bare}(\mathrm{de\
Sitter}) \sim e^{10^{122.44}}$.  This is the same type of argument
that was used in \cite{Haw-Page} to say that the Hartle-Hawking
no-boundary proposal leads to the prediction that the flatness
parameter has unit probability to be $\Omega=1$ (in a minisuperspace
model that did not include inhomogeneous modes that would
realistically be expected to give some cosmic variance about
$\Omega=1$ in the observed part of the universe), despite the fact
that the no-boundary wavefunction peaks at the minimum possible value
of inflation.

	Although we did not refer to the tunneling wavefunction in
this paper \cite{Haw-Page} on the flatness of the universe, our
argument for the infinite measure from the integral over $\phi_0$
would also answer the challenge of Vilenkin \cite{VvsHH} and others
\cite{BC,GrT,BP} that the no-boundary proposal favors a small amount
of inflation, whereas the tunneling wavefunction favors a large
amount.  Our argument would imply that even if there is a huge bare
probability (i.e., before normalization) for a small amount of
inflation in the no-boundary proposal, if one includes an infinite
range for the initial value $\phi_0$ of the inflaton field, that
gives an infinite measure, which of course dominates over the large
but finite measure or bare probability for a small amount of
inflation.

	In the case of Susskind's challenge from the huge negative
action of a large Euclidean de Sitter solution, the problem is
quantitatively more acute, since the action of a large Euclidean de
Sitter solution is even enormously much more negative than that of the
smallest theoretical amount of inflation ($\phi_0 \sim 1$, much
smaller than the observational lower limit for at least roughly 50
e-folds of inflation).  However, this huge negative action is still
finite, so qualitatively the solution of an infinite range for
$\phi_0$ can still work just as it did in the previous case (assuming
that it did there).

	It is amusing to consider the quantitative implications of
this proposed resolution of Susskind's challenge.  If one imagined
that $\phi$ really has only a finite range but attempted to make that
range so large, say up to $\phi_{\mathrm{max}}$, that the integral
over $d\phi_0$ dominates over $P_{\rm bare}(\mathrm{de\ Sitter}) \sim
e^{10^{122.44}}$, we would need $\phi_{\mathrm{max}}
\stackrel{>}{\sim} e^{10^{122.44}}$.  Then if $V(\phi)$ rises
asymptotically as some power of $\phi$, the amount of inflation as
$\phi$ undergoes slow roll from near $\phi_{\mathrm{max}}$ to near
unity (at the end of inflation) is exponential in a power of
$\phi_{\mathrm{max}}$.  If this power is an exponent that is of the
order of magnitude of unity, then after inflation, the size of the
universe will be at least of the crude order of $10^{10^{10^{122}}}$
(meaning that the logarithm of the logarithm of the logarithm of the
size will be at least roughly 122).  Of course, it is hard to imagine
why $\phi$ would have a finite range if its range extended up to at
least roughly $10^{10^{122}}$, so it seems much more likely that
$\phi$ would then simply have an infinite range.  In that case, the
probability would then be unity that the universe would expand larger
than any fixed finite size.  This is effectively almost the same as
saying that the universe will have an infinite amount of expansion,
though strictly speaking ``infinite'' here should be taken to mean
just ``arbitrarily large.''

	The main problem with this proposed resolution of Susskind's
challenge is that it generally requires that the inflaton be allowed
to be so large that its potential gives energy densities far in
excess of the Planck density (unless the potential levels off below
the Planck value, which is a distinct possibility, though perhaps one
that would be considered to be fine tuned).  Then one might suppose
that the inflaton should be cut off at the Planck density.  However,
even while admitting that we do not yet know what should happen at
the Planck density, one might say \cite{Page-space} that this cut off
is {\it ad hoc}, so we cannot be sure that the proposed solution,
with $\phi$ allowed to be infinitely large (or at least as large as
$P_{\rm bare}(\mathrm{de\ Sitter}) \sim e^{10^{122.44}}$), is not
qualitatively valid.  In other words, even though it may be doubtful
that it is really right, we cannot be sure it is wrong either.

	A related problem is that if the inflaton comes from some
modulus or other field in superstring/M theory,  there are
conjectures that in the presence of gravity the volume of the moduli
space is finite (see, e.g., \cite{OV}).  If so, the integral over the
allowed range of the inflaton field would give a finite answer that
almost certainly not compensate for $P_{\rm bare}(\mathrm{de\
Sitter})$.  This means that it may be hard to combine this proposed
solution to Susskind's challenge with superstring/M theory.  However,
this is just an unproved conjecture \cite{Lpriv}, and the KKLT
construction \cite{KKLT} suggests that it may well be wrong.

	(8) An eighth possible resolution of Susskind's challenge is
that the inflationary component of the wavefunction expands to such
an utterly enormous size that it produces more ordered observers than
the nearly empty de Sitter spacetime does of disordered observers
through vacuum fluctuations, even when one includes the huge bare
probability of the nearly empty de Sitter spacetime.  For this
resolution to work, one would need to restrict the 4-volume of the de
Sitter spacetime (e.g., by something that prevented it from lasting
too long and expanding too many times, perhaps the same thing that
might prevent too many observers from occurring by vacuum
fluctuations in the future of our subuniverse
\cite{Page-lifetime,decay,BF,LindeBB,return,Vilfreak,astronomical})
so that it produces a strictly finite number of observers, and then
allow the inflationary universe to expand so much more that it
produces more observers, even after including the ratio of their bare
probabilities that seems to weight the nearly-empty de Sitter
spacetime by such a large factor relative to the inflationary
solution.

	In the case of a minisuperspace comparison between the
tunneling and the no-boundary quantum states \cite{Page-space}, for
suitable potentials (including the simple massive scalar inflaton),
even deterministic slow roll without stochastic inflation can produce
enough volume from a large enough $\phi_0$ to compensate for the
higher bare probability of a small $\phi_0$ (and a resulting amount
of inflation too small to be consistent with observations), without
having to go to $\phi_0$ so high that one exceeds the Planck
density.  However, for the same minisuperspace idea to save the
no-boundary proposal in comparison with the Euclidean de Sitter
extremum, it appears that for most reasonable potentials that rise
indefinitely with the inflaton field, one would need to allow the
initial energy density to exceed the Planck value.

	However, if one goes to stochastic or eternal inflation
\cite{Veternal,Leternal,Seternal,Geternal}, it appears to allow the
universe to inflate to arbitrarily large size even without the
potential ever exceeding the Planck value (though for an ever-rising
potential it does seem that the stochastic evolution pictured for the
scalar field would be required to be rather finely tuned to avoid
ever exceeding the Planck energy density; this problem would not
arise if the potential instead has a maximum value below the Planck
value \cite{Vpriv}).  In this case one could always get enough
spatial volume, and hence number of observers, in the inflationary
solution to compensate for the enormous unnormalized bare probability
of the Euclidean de Sitter spacetime, assuming that the latter
somehow is not similarly allowed to inflate by a sufficient amount to
produce its own larger number of observers by vacuum fluctuations.

	One might think that including the processes of stochastic
inflation would take one outside the zero-loop approximation
advocated in \cite{Page-Hawking-BD} to avoid some of the infinities
of the path integral.  However, one might conjecture that the effects
of stochastic inflation could arise from taking into account complex
inhomogeneous classical solutions of the field equations (extrema of
the action).  It would be very interesting to see whether this indeed
is the case.

	Another way to get an arbitrarily large amount of inflation
is to suppose that the inflaton potential has a rather flat maximum,
and that inflation starts at the top of this hill \cite{V,TL,HH1}. 
In this case one could get homogeneous complex classical extrema with
arbitrarily large amounts of approximately real Lorentzian inflation,
expanding the universe to arbitrarily large size.  However, one could
object that the de Sitter-like extrema corresponding to the currently
observed cosmic acceleration can also expand the universe to
arbitrarily large size in the distant future, so it is not obvious
why the arbitrarily large size from rapid early inflation would
dominate over the arbitrarily large size from the slow late
inflation.

	Again it seems that we must imagine that for some reason the
large nearly-empty de Sitter solution cannot do something that the
rapid-inflation solution can.  In the first proposed resolution
above, it was the formation of observers that was proposed to be
denied the nearly-empty de Sitter solution.  In the second suggestion
it was supposed that the nearly-empty de Sitter solution cannot be
attached by a bridge to another large 4-sphere to make its Euclidean
action enormously more negative as it was proposed could happen to
the inflationary solution.  In the third speculation, it was supposed
that the nearly-empty de Sitter solution cannot be combined with
another space to give an actual extremum with greatly reduced action,
even though it was conjectured that this might be able to be done for
the inflationary solution.  In the fourth suggestion, it was proposed
that Euclidean de Sitter is not actually an allowed extremum for the
Hartle-Hawking path integral, whereas the inflationary solution
supposedly is.  In the fifth idea, it was suggested that Euclidean de
Sitter might not be a solution at all for whatever it actually is
that is driving the currently observed cosmic acceleration.  In the
more vague sixth proposal, the solution to the infinite measure
problem is supposed to reduce the na\"{\i}ve de Sitter bare
probability much more than that of inflation.  In the seventh
proposed solution to Susskind's challenge, it is the arbitrarily
large range of $\phi_0$ that the inflationary solutions have that the
de Sitter solution does not have.  (Here it perhaps is most easy to
see the distinction between the two solutions, which is why I am
perhaps most attracted to this possibility.)  Finally, in the eighth
possibility, it is proposed that the de Sitter space cannot expand
large enough to produce arbitrarily many observers (by vacuum
fluctuations), even though the inflationary universe can (though in
this case by the ordinary evolutionary process that we believe
occurred in our observed subuniverse).

	Thus we have at least an eight-fold way of potential
solutions to save the Hartle-Hawking no-boundary proposal (and what
it might explain, such as the mysterious arrow of time) from
Susskind's challenge.  As one can see from the discussion above, I am
not too happy with any of them, but at the moment I would guess that
the seventh, with the infinite measure from the integration over an
infinite range of possible initial values of the inflaton scalar
field $\phi$, seems the least unattractive.

\section*{Other Possibilities}

	Since it is not certain whether any of these eight proposals
(or others I have not yet thought of or that other people might
propose) really give a satisfactory resolution of Susskind's
challenge, let us now turn to the possibility that the Hartle-Hawking
no-boundary proposal is wrong and that one should turn to another
proposal for the quantum state of the universe.  Here I shall just
examine the tunneling proposals of Vilenkin, Linde, and others
\cite{V,TL,TZS,TR,TVV,TGV}.

	For the present purposes, the main difference from the
Hartle-Hawking proposal will be taken to be the sign of the Euclidean
action for at least the homogeneous isotropic complex Euclidean FRW
solutions like Euclidean de Sitter and FRW inflation \cite{TL,V}. 
(It seems problematic to take the opposite sign for inhomogeneous
and/or anisotropic perturbations without leading to some
instabilities, and it is not clear how to give a sharp distinction
between the modes that are supposed to have the reversed sign of the
action and the modes that are supposed to retain the usual sign of
the action, but for this paper I shall generally leave aside this and
related problems \cite{BH,GV97,Lin98,HT,TH,V98}.  Vilenkin has
emphasized \cite{V} that this criticism does not seem to apply to his
tunneling proposal, which does not simply have the reversed sign of
the Euclidean action for all modes, but here I shall just focus on
the homogeneous mode, for which Vilenkin's proposal effectively does
have the opposite sign.)

	In this case the Euclidean de Sitter solution would give
\begin{equation}
P_{\rm bare}(\mathrm{de\ Sitter}) = e^{+2S_E(\mathrm{dS})}
 = e^{-\pi a_{\mathrm{dS}}^2} 
 = e^{-3\pi/\Lambda} \sim e^{-10^{122.44}}.
\label{eq:11}
\end{equation}
Assuming that a vacuum fluctuation producing an observer has the usual
sign of the Euclidean action, one would then get
\begin{eqnarray}
P_{\rm observation, unnormalized}(\mathrm{de\ Sitter}) \sim
\exp{(-10^{122.44} - 10^{69.4})}.
\label{eq:12}
\end{eqnarray}

	These bare probabilities could then be compared with the
inflationary probabilities
\begin{equation}
P_{\rm bare}(\mathrm{inflation}) = e^{+2S_E(\mathrm{inflation})}
 = e^{-\pi a_0^2} = e^{-3/[8V(\phi_0)]}
 \stackrel{>}{\sim} e^{-10^{10.6}}
\label{eq:13}
\end{equation}
and
\begin{eqnarray}
P_{\rm observation, unnormalized}(\mathrm{inflation})
\stackrel{>}{\sim} e^{-10^{10.6}}.
\label{eq:14}
\end{eqnarray}

	This dominates the corresponding $P_{\mathrm observation,
unnormalized}({\mathrm{de\ Sitter}})$, so if we again normalize be
dividing by the total unnormalized probability for observations, for
the tunneling wavefunction we now get that the normalized probability
for an observation to occur in an inflationary solution would be
\begin{eqnarray}
P_{\rm observation}(\mathrm{inflation}) \sim 1.
\label{eq:15}
\end{eqnarray}
Thus the tunneling wavefunction would be consistent with our ordered
observations in this way (at least if one could solve the other
problems associated with it).

	It is a bit disconcerting that the controversy between the
no-boundary and tunneling wavefunctions
\cite{BH,GV97,Lin98,HT,TH,V98} has not yet been resolved.  In terms
of the numbers above, they give probabilities of large empty de
Sitter spacetimes that differ by a factor of more than
$10^{10^{122}}$, which is the ten thousand million million millionth
power of a googolplex!  However, even this might pale beside the
uncertainties of whether the various infinite factors discussed above
should be included (particularly that of the integration over an
infinite range of the initial value $\phi_0$ of the inflaton).

	One argument \cite{Veternal,Leternal,Seternal,Geternal,LLM}
is that at very late times, where the volume of space has grown so
large that that is where almost all observers are expected to be,
eternal inflation leads to the same predictions for all of the
various proposed wavefunctions.  This picture is now being explored
in the context of the string landscape \cite{BP,KKLT,Sus03,BDG,
FST,FS,FKMS,Vafa, Susskind-book,FSSY,Villand,BFL,VV,OV}, with one of
the recent ideas being that the probabilities of the various string
vacua depends not only on the various actions but also on the decay
rates \cite{Bousso,CDGKL,LindeBB}.  Whether these ideas can be cast
into the form of the Hartle-Hawking no-boundary proposal or are
consistent with it remains to be fully explored.  On the other hand,
the canonical classical measure \cite{Hen,GHS} gives an ambiguous
probability for inflation \cite{HP88}.  Gibbons and Turok have
recently shown \cite{GT} that the divergence in the canonical measure
is removed if one identifies universes which are so flat they cannot
be observationally distinguished by observers like us, living in the
late universe but with access to only a finite portion of space in
the past. With this identification of very flat universes, the
canonical measure becomes finite, and it gives an exponentially small
probability for a large number of inflationary e-foldings.  However,
\cite{HP88} also implies that one can alternatively choose other
cutoffs in which the probability of inflation is large
\cite{Page06}.  These examples show that it is certainly not the case
that all choices of measure (or initial conditions or wavefunctions)
lead to the same predictions, so one would really like to know what
the quantum state is and what measure it predicts for observations.

	In summary, Susskind has raised a serious challenge to the
Hartle-Hawking no-boundary proposal for the quantum state of the
universe.  There are several potential resolutions of this challenge,
but it is not yet clear whether any of them is satisfactory.  If no
resolutions can be found, the challenge leaves us with the mystery of
what the quantum state might be to be consistent with our
observations.

\newpage

\section*{Acknowledgments}

	I appreciate Lenny Susskind's explaining to me on several
occasions his objections to the Hartle-Hawking no-boundary proposal. 
I am also grateful for the hospitality of the University of British
Columbia during and after the Unruh-Wald fest, where I had key
discussions on Susskind's challenge with Jim Hartle and Bill Unruh. 
Andrei Linde and Alex Vilenkin provided many useful comments on the
manuscript and on eternal inflation and the string landscape.  Other
helpful email comments were given by Andreas Albrecht, Anthony
Aguirre, Nick Bostrom, Raphael Bousso, Bernard Carr, Sean Carroll,
David Coule, William Lane Craig, George Ellis, Gary Gibbons, Steve
Giddings, J. Richard Gott, Jim Hartle, Pavel Krtou\v{s}, John Leslie,
Don Marolf, Joe Polchinski, Martin Rees, Michael Salem, Mark
Srednicki, Glenn Starkman, Lenny Susskind, and Neil Turok.  This
research was supported in part by the Natural Sciences and
Engineering Research Council of Canada.

\end{document}